\documentclass[preprint]{aastex}
\pdfoutput=1
\usepackage{url}
\usepackage{afterpage}
\slugcomment{Accepted to the \apj}
\shorttitle{SED Fitting of HPS LAEs}
\shortauthors{Hagen et al.}
\newcommand{\eg}{e.g.,}

\newcommand{\etal}{et~al.}
\begin{document}

\title{Spectral Energy Distribution Fitting of HETDEX Pilot Survey 
Ly$\alpha$ Emitters in COSMOS and GOODS-N}

\author{Alex Hagen\altaffilmark{1,2}, Robin Ciardullo\altaffilmark{1,2}, 
Caryl Gronwall\altaffilmark{1,2}, Viviana Acquaviva\altaffilmark{3}, 
Joanna Bridge\altaffilmark{1,2}, Gregory R. Zeimann\altaffilmark{1,2},
 Guillermo A. Blanc\altaffilmark{4}, Nicholas A. Bond\altaffilmark{5}, 
 Steven L. Finkelstein\altaffilmark{6}, Mimi Song\altaffilmark{6}, 
 Eric Gawiser\altaffilmark{7}, Derek B. Fox\altaffilmark{1,2}, Henry Gebhardt\altaffilmark{1,2}, 
A.I. Malz\altaffilmark{1,2}, Donald P. Schneider\altaffilmark{1,2}, 
\author{Niv Drory\altaffilmark{8}, Karl Gebhardt\altaffilmark{8}, Gary J. Hill\altaffilmark{8}}}
\email{hagen@psu.edu}

\altaffiltext{1}{Department of Astronomy \& Astrophysics, The Pennsylvania State 
University, 525 Davey Lab, University Park, PA 16802}
\altaffiltext{2}{Institute for Gravitation and the Cosmos, 
The Pennsylvania State University, University Park, PA 16802}
\altaffiltext{3}{Department of Physics, New York City College of Technology, 
City University of New York, 300 Jay Street, Brooklyn, NY 11201}
\altaffiltext{4}{Observatories of the Carnegie Institution for Science, 
Pasadena, CA 91101}
\altaffiltext{5}{Cosmology Laboratory, NASA Goddard Space Flight Center, 
Greenbelt, MD 20771}
\altaffiltext{6}{Department of Astronomy, The University of Texas at Austin, 
Austin, TX, 78712}
\altaffiltext{7}{Department of Physics and Astronomy, Rutgers, 
The State University of New Jersey, Piscataway, NJ 08854}
\altaffiltext{8}{Department of Astronomy, The University of Texas at Austin, 
Austin, TX, 78712}

\begin{abstract}

We use broadband photometry extending from the rest-frame UV to the
near-IR to fit the {\it individual\/} spectral energy distributions (SEDs)
of 63 bright ($L({\rm Ly}\alpha) > 10^{43}$~ergs~s$^{-1}$) 
Ly$\alpha$ emitting galaxies (LAEs) in the redshift range $1.9 < z < 3.6$.
We find that these LAEs are quite heterogeneous, with stellar masses that span 
over three orders of magnitude, from $7.5 < \log M/M_{\odot} < 10.5$.   Moreover,
although most LAEs have small amounts of extinction, some high-mass objects have 
stellar reddenings as large as $E(B-V) \sim 0.4$.  Interestingly, in
dusty objects the optical depths for Ly$\alpha$ and the UV continuum
are always similar, indicating that Ly$\alpha$ photons are not undergoing
many scatters before escaping their galaxy.  In contrast, the ratio of 
optical depths in low-reddening systems can vary widely, illustrating the
diverse nature of the systems.  Finally, we show that in the 
star formation rate (SFR)-log mass diagram, our LAEs fall above the
``main-sequence'' defined by $z \sim 3$ continuum selected star-forming galaxies.
In this respect, they are similar to sub-mm-selected galaxies, although
most LAEs have much lower mass.

\end{abstract}

\section{Introduction}
\label{sec:intro}

\cite{par67} originally predicted that the Ly$\alpha$ emission line could be 
a very useful probe of the high-redshift universe, and, while it took many 
years to detect this feature \citep{cow98, hu98}, Ly$\alpha$ emitting 
galaxies (LAEs) are now routinely observable from $z \sim 0.2$ \citep{deh08, 
cow10} to $z>7$ \citep{hu10, ouc10, lid12, ono12}.  
However, while the detection of Ly$\alpha$ in the high-redshift universe is 
relatively common, the physics of this emission is still not well understood. 
Since Ly$\alpha$ is a resonance transition, it is likely that each
photon scatters many times off intervening neutral material before 
escaping into intergalactic space.   As a result, even a small amount of dust 
should extinguish the line, and indeed, only $\sim 25$\% of 
Lyman-break galaxies (LBGs) at $z\sim2-3$ have enough
Ly$\alpha$ in emission to be classified as an LAE  \citep{sha03}.   
While it is possible for dusty galaxies to
create an escape path for Ly$\alpha$ via supernova-blown bubbles and/or 
exotic geometry \citep[\eg][]{ver12}  
most analyses suggest that the LAE population as a whole is made up of 
young, low-mass, low-metallicity systems, possessing relatively little 
interstellar dust \citep[\eg][]{gaw07, gua11}.   

To date, most Ly$\alpha$ emitters have been detected via deep narrow-band 
imaging with 4-m and 8-m class telescopes \citep[\eg][]{gro07, ouc08} .  
These surveys generally extend to low Ly$\alpha$ luminosities and sample a 
wide range of the high-redshift galaxy luminosity
function.  Unfortunately, in the continuum, LAEs are usually quite faint, 
which makes studying their spectral energy distributions (SEDs) 
difficult.  As a result, most of our knowledge about those physical properties which
are encoded in the objects' SEDs -- information such as stellar mass, extinction, and
population age -- has come from stacking techniques \citep[\eg][]{gaw07, gua11}.  These analyses 
only yield estimates for a ``typical'' LAE and may be subject to serious systematic biases associated 
with the stacking techniques \citep{var13}. Moreover, those few programs that have
sought to measure the SEDs of individual LAEs \citep[\eg][]{fin09, nil09, yum10, nak12, mcl14} have generally been 
restricted to very small numbers of objects.   These efforts have been able to provide hints as to the 
range of properties exhibited by the population, but have been unable 
to probe the statistics of the entire LAE population. Thus, while we have some idea about the mass and 
dust content of ``representative" LAEs, the distribution of physical parameters 
for the entire population remains poorly constrained. 

Here, we investigate the stellar populations of luminous Ly$\alpha$ emitters
by analyzing the individual spectral energy distributions of 63 
$1.9 < z < 3.6$ LAEs detected by the McDonald 2.7-m telescope's 
Hobby-Eberly Telescope Dark Energy Experiment (HETDEX) Pilot Survey.
In Section~\ref{sec:sample}, we summarize the HETDEX Pilot Survey and describe 
the ancillary groundbased, {\sl HST,} and {\it Spitzer\/} photometry which is 
available for analysis.   In Section~\ref{sec:analysis}, we briefly describe 
the SED-fitting code {\tt GalMC} \citep{acq11} and the underlying assumptions 
used to derive stellar mass, extinction, and age from a set of broadband 
photometry which extends from the rest-frame UV through to the near-IR\null.  
We also outline the procedures used to measure the physical sizes of the LAEs in a 
manner that is insensitive to the effects of cosmological surface brightness 
dimming.  In Section~\ref{sec:results}, we present our results and show that 
the population of luminous $z \sim 3$ LAEs is quite heterogeneous, with sizes 
extending from $0.5$~kpc $ \lesssim r \lesssim 4$~kpc, stellar masses ranging from 
$7.5 < \log M/M_{\odot} < 10.5$, and differential extinctions varying between 
$0.0 < E(B-V) < 0.4$.  We illustrate several trends involving LAE physical 
parameters, including a positive correlation between reddening and stellar 
mass,  a positive correlation between stellar mass and galactic age, and a 
positive correlation between galaxy size and Ly$\alpha$ luminosity.  We also 
examine the possible evolution of physical properties with redshift and compare 
our LAEs to other $z \sim 3$ objects on the star-forming galaxy main sequence. 
We conclude by discussing the implications of our results for the underlying physical 
mechanisms of Ly$\alpha$ escape in high redshift galaxies.  

For this paper we adopt a cosmology with $H_0 = 70$~km~s$^{-1}$~Mpc$^{-1}$, 
$\Omega_{\rm M} = 0.3$, and $\Omega_\Lambda = 0.7$ \citep{pla13, hin12}.

\section{Our Sample} 
\label{sec:sample}
The LAEs chosen for study were  discovered with the George \& Cynthia Mitchell 
Spectrograph \citep[previously known as VIRUS-P;][]{hil08b} on the 2.7-m 
Harlan J. Smith Telescope during the HETDEX Pilot Survey \citep[HPS;][]{ada11}.
This integral-field instrument, which employs an array of 246 $4\farcs 2$ 
diameter fibers, covers $\sim 3$~arcmin$^{2}$ of sky at a time, and delivers
5~\AA\ resolution spectra between the wavelengths 3500~\AA\ and 5800~\AA\null.
The HPS itself surveyed a total of 169~arcmin$^2$ in the COSMOS \citep{sco07}, 
GOODS-N \citep{gia04}, MUNICS-S2 \citep{dro01}, and 
XMM-LSS \citep{pie04} fields and reached a limiting line flux of 
$6.7 \times 10^{-17}$~ergs~cm$^{-2}$~s$^{-1}$ at 5000~\AA\ (for 50\% of its 
pointings) and $1.0 \times 10^{-16}$~ergs~cm$^{-2}$~s$^{-1}$ at 
5000~\AA\ (for 90\% of the pointings).  The final HPS catalog consists of 
coordinates, redshifts, $R$-band magnitudes, line fluxes, and equivalent 
widths for 397 emission-line selected galaxies.  Ninty-nine of these 
sources are non X-ray emitting LAEs with $1.9 < z < 3.8$, rest-frame 
equivalent widths EW$_0 > 20$~\AA, and Ly$\alpha$ luminosities between
$2.6 \times 10^{42}$ and $1.1 \times 10^{44}$~ergs~s$^{-1}$ \citep{ada11, 
bla11}.   A total of 74 of these HPS LAEs lie in the GOODS-N and COSMOS 
fields, where deep {\sl HST\/} data is available. The redshift range for this subsample
is $1.9 < z < 3.6$.

The process of assigning an optical counterpart to each HPS emission-line 
detection was challenging.  As pointed out by \citet{ada11}, there
is an order of magnitude mismatch between the spatial resolution obtained
from the $4\farcs 2$ diameter fibers of VIRUS-P, and that delivered by
the broadband imagers of {\sl HST.}  Thus, each assignment was done in a
Bayesian manner, by calculating the likelihood of association for each
object within a $10$\arcsec\ window of the nominal position obtained from
the spectroscopy \citep[see Section 5.3 of][]{ada11}. Formally, the median 
probability for identifying the correct optical counterpart was 64\%. However, as discussed 
below in the final paragraph of Section \ref{sec:sed}, there is no statistical difference between 
the distribution of SED properties for a sample LAEs with high-probability and/or confirmed 
counterparts and that for the sample of lower-probability associations.

In total, we identified 67 HPS LAEs with optical counterparts. Four of these
objects (HPS IDs 144, 145, 160, and 196) were removed from our analysis based on the work of 
\cite{bla11}, who showed that their UV slopes were more consistent with those of 
foreground [\ion{O}{2}] emitters than LAEs.  This culling left us with 63 objects for analysis.  Since
the X-ray data in GOODS-N and COSMOS is deep enough to rule out most AGN,
we believe that the bulk \textbf{of} these objects are true Ly$\alpha$ emitting
sources with $1.9 < z < 3.6$ and monochromatic Ly$\alpha$ luminosities between 
$3.4 \times 10^{42}$ and $3.8 \times 10^{43}$~ergs~s$^{-1}$.

Before proceeding further, we should note that the LAEs discovered by the 
HPS are significantly more luminous than the Ly$\alpha$ emitters found by 
most narrow-band surveys.  While the $2 < z < 3$ observations of 
\citet{gro07}, \citet{ouc08}, and \citet{gua10} typically reach 
Ly$\alpha$ 90\% completeness levels of $L({\rm Ly}\alpha) \sim 
10^{42}$~ergs~s$^{-1}$, the median HPS limit is five times brighter than 
this.  On the other hand, since the Ly$\alpha$ luminosity limit of the HPS is 
very nearly constant across the survey's entire spectral range \citep[see 
Figure~1 of][]{bla11}, the data set covers an order of magnitude 
more co-moving volume than a typical narrow-band survey, with 
$V = 5.63 \times 10^5$~Mpc$^{3}$ in the COSMOS and GOODS-N regions alone.  
This allows us to obtain good statistics on the bright end of the LAE 
population, and explore evolution over $\sim 1.6$~Gyr of 
cosmic time.

The HPS fields, and in particular, the COSMOS and GOODS-N regions are rich in 
deep, ancillary imaging and provide up to 18 photometric data points for SED fitting. Tables~\ref{table:cosmosphot} and
\ref{table:goodsnphot} summarize these data.  Most of the HPS/COSMOS and 
HPS/GOODS-N fields are part of CANDELS 
\citep{grog11, koe11}, and 49 out of our 63 LAEs have deep 
{\sl HST\/} optical and near-IR photometry \citep{son13, fin13} from
this program.  Moreover, all of our targets have
photometry from \citet{ada11}, who used the deep ground-based images of
COSMOS as their source frames \citep{cap07}. Note that many of the LAEs 
targeted in this survey are too faint in the continuum to be present 
in the published COSMOS and GOODS photometric catalogs; for these objects,  
AB magnitudes were determined by re-measuring the original images using the 
positions of the {\sl HST\/} optical counterparts. Still, there are some non-detections. 
When this occurred, an upper flux limit was assigned as the $1 \, \sigma$ uncertainty
of the local sky value.   In some cases, these limits were crucial for constraining
the SED properties of our targets.

Data at longer wavelengths come from observations with the Spitzer telescope.  
Once again, most LAEs are far too faint to be present in the S-COSMOS and 
GOODS-N Spitzer catalogs, as these analyses have relatively high detection 
thresholds (1 $\mu$Jy in IRAC channel 1).  Since the rest-frame near-IR is 
extremely important for determining stellar mass, we performed our own
aperture photometry on these frames using MOPEX\footnotemark \footnotetext{Information on MOPEX is available at \url{http://irsa.ipac.caltech.edu/data/SPITZER/docs/dataanalysistools/tools/mopex/}} \citep{mak05} at the known 
LAE positions.  After experimenting with a variety of apertures, we settled 
on a photometric radius of $3\farcs 6$, and then applied an aperture 
correction as described in the IRAC Instrument 
Handbook\footnotemark\footnotetext{http://irsa.ipac.caltech.edu/data/SPITZER/docs/irac/iracinstrumenthandbook/28/}. 

\section{Data Analysis}
\label{sec:analysis}
\subsection{Spectral Energy Distribution Fitting}
\label{sec:sed}
The spectral energy distribution of a galaxy encodes a number of physical 
parameters, including stellar mass, age, dust content, and the current star 
formation rate (SFR\null).  For example, since a galaxy's near-IR flux arises 
principally from the evolved stars of all stellar populations, that part of 
the SED traces the system's total stellar mass \citep{bel01, zib09}.  In 
contrast, the slope of a galaxy's far UV ($\sim 1600$~\AA) continuum is 
fixed by the Rayleigh-Jeans tail of the blackbody emission from hot, young 
stars. The amplitude of the UV continuum thus yields the star formation rate and 
any flattening of the UV continuum's slope is most likely due to the effects of 
dust \citep{ken98, cal01}.  Estimates of population age come primarily from the regions in 
between, as features such as the Balmer and 4000~\AA\ breaks are sensitive to 
the main sequence turnoff and the exact mix of
intermediate age stars \citep[\eg][]{kau03}.  

To extract this information, we began with the population synthesis models 
of \citet{bru03}, which were updated in 2007 (BC07) with an improved treatment 
of the thermal-pulsing asymptotic giant branch (TP-AGB) phase of stellar evolution.
This phase of stellar evolution can be important for systems older than
$\sim10^8$ years, which is $\sim 30\%$ of our sample (see Section~\ref{sec:age}).
We also performed fits using the older BC03 models, but due to the generally young
ages of the stellar populations, these fits were statistically indistinguishable from the 2007 
models.  For the remainder of this paper, we will only refer to our BC07 results.
For consistency with the works of \citet{gua11} and \citet{acq11}, we adopted 
a \citet{sal55} initial mass function (IMF) over the range 
$0.1 M_{\odot} < M < 100 \, M_{\odot}$, a \citet{cal01} extinction law, and a 
\citet{mad95} model for the effects of intervening intergalactic absorption. 
Since stellar metal abundances are poorly constrained by broadband SED 
measurements, we fixed the metallicity of our models to $Z = 0.2 \, Z_{\odot}$;
this is roughly the gas-phase abundance inferred from recent near-IR 
spectroscopy of $z \sim 2$ LAEs \citep{nak13, son13}.  Emission lines and nebular continuum, 
which can be an important contributor to the broadband SED of high-$z$ galaxies 
\citep[\eg][]{ate11, sch09}, were modeled following the prescription of \citet{acq11} 
with updated templates from \cite{acq12}.  Finally, following \citet{gua11}, 
we adopted the simple assumption that the SFRs of our LAEs have been constant
with time.

In keeping with these assumptions, we did not use any of the bandpasses 
listed in Tables~\ref{table:cosmosphot} and \ref{table:goodsnphot}
that lie redward of $3.3~\mu$m in the rest-frame.   This is where the first
PAH line is located, and such ISM features are not accounted for in the stellar 
populations models.  We also did not use data bluewards of Ly$\alpha$, as 
the \citet{mad95} correction is statistical in nature, and large excursions
from the norm could bias our reddening measurements.

Since SED fitting is a notoriously non-linear problem that may involve many 
local minima, highly non-Gaussian errors, and {degeneracies between parameters}, 
we chose to analyze our data using {\tt GalMC}, a Markov-Chain Monte-Carlo (MCMC) code 
with a Metropolis-Hastings sampler \citep{acq11}.  This approach is much more 
computationally efficient than traditional grid searches, as it explores
all regions of parameter space, while still spending the bulk of its time in the 
highest-likelihood parts of the probability distribution. The uncertainties associated with the
fitted parameters are also much more realistic than those estimated using a simple $\chi^2$ minimization,
as degeneracies between the variables are bettered explored and quantified.
For each SED,  three free parameters, stellar mass, $E(B-V)$, and age (which, under the
assumption of a constant SFR history, is equivalent to star formation rate),
were fit using four chains initiated from random starting locations.  Once completed, the chains 
were analyzed via the {\tt CosmoMC} program {\tt GetDist} \citep{lew02}, and, 
since multiple chains were computed for each object, the \citet{gel92} $R$ 
statistic was used to test for convergence using the criterion $R - 1 < 0.1$
\citep{bro98}.

As discussed at length by \citet{con13}, the results of our fits should be 
robust within the context of our model assumptions.  Of course, any change to 
these assumptions will result in a systematic error.   For example, the use of 
a \citet{cha03} or \citet{kro01} IMF would systematically reduce our stellar 
mass estimates by $\sim 0.3$~dex,  while leaving our values for extinction and age 
essentially unchanged \citep{pap11}.  Similarly, a different treatment
of the TP-AGB phase may change the stellar mass estimates by up to $\sim 0.3$~dex 
\citep{zib09}, while a shift to solar metallicity will generally increase 
our masses by $\sim 0.1$~dex.  A full discussion of these systematic 
uncertainties is given by \citet{con13}.  

\begin{figure}[t]
\plotone{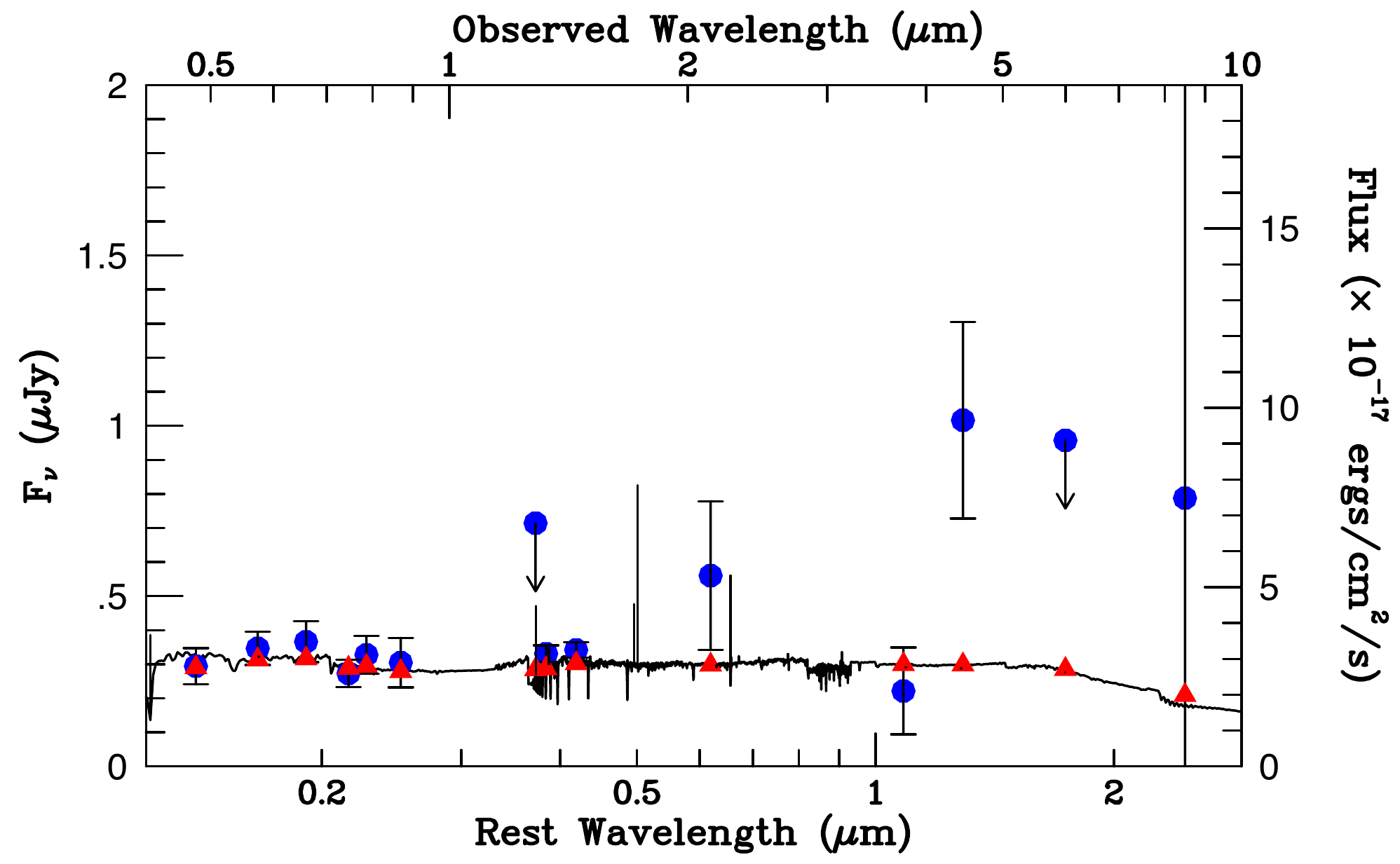}
\caption{The results of our SED fit to the photometry of HPS189. The blue points are the
observed flux densities, the black line is the best fit SED, and the red triangles show
the predicted flux density within each band.  The left axis defines the scaling for the
galactic continuum; the right axis gives the monochromatic flux scale applicable to the
emission lines.}
\label{fig:hps288}
\end{figure}

Figure~\ref{fig:hps288} demonstrates the effectiveness of our fits by comparing 
the broadband photometry to the best fit SED for HPS 189. Since our data contain
many bands of photometry, the SEDs of our galaxies are generally well constrained.
Table~\ref{table:sed} summarizes the SED-based properties of all 63 LAEs in 
our sample.

\begin{figure}[t]
\centering
\includegraphics[scale=0.8]{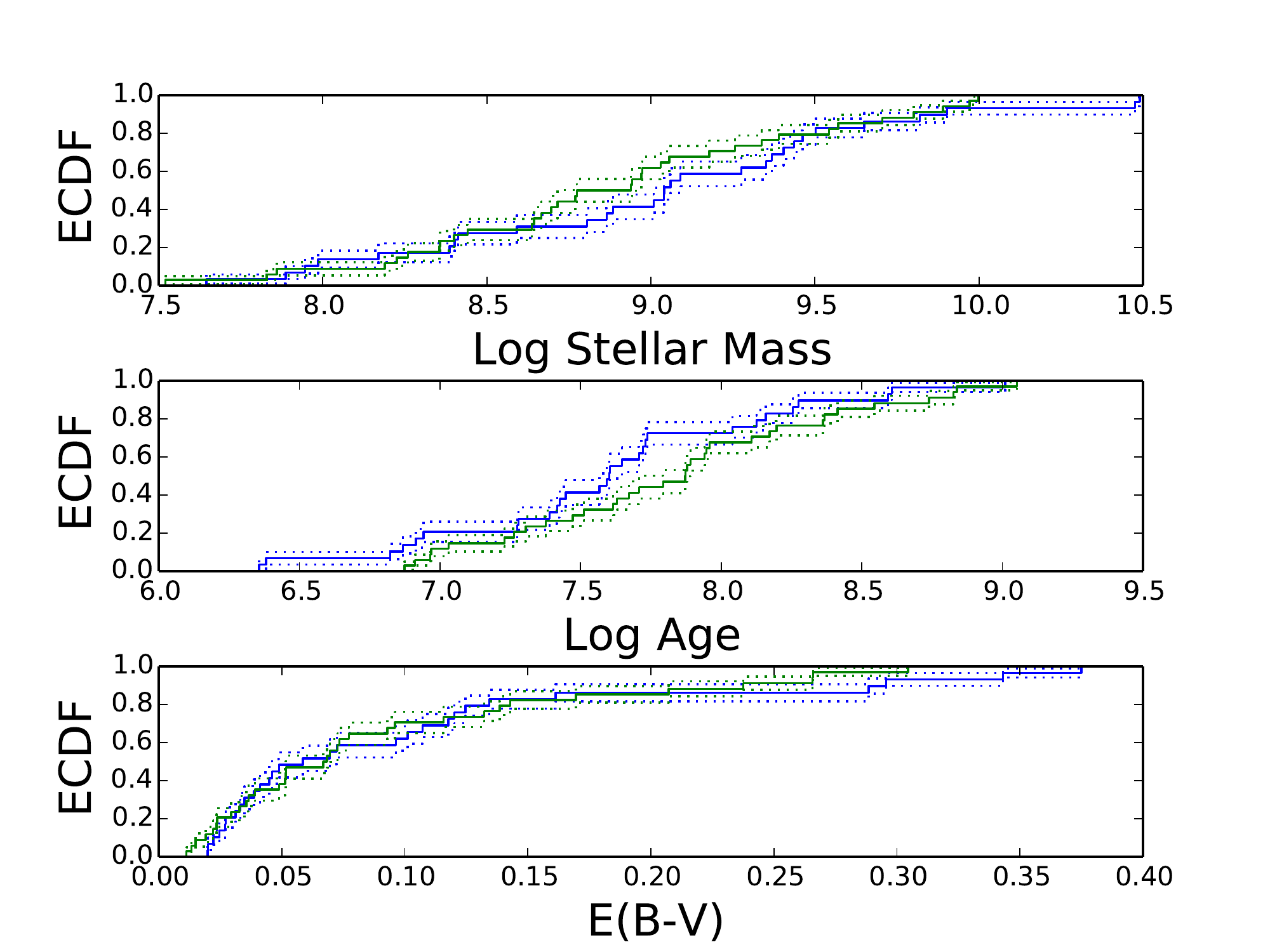}
\caption{The Empirical Cumulative Distribution Functions of the clean sample of LAEs (blue), compared to
those for LAEs with lower-probability counterparts (green).  The dotted lines represent the $1 \, \sigma$
asymtotic errors \citep{don52}.  In all cases the K-S test cannot reject the null hypothesis that both 
samples are drawn from the same underlying distribution.}
\label{fig:samples}
\end{figure}

As stated in Section \ref{sec:sample}, the assignment of optical counterparts
to the HPS-discovered emission lines is probabilistic in nature.  To
investigate this further, we constructed a ``clean'' sample of HPS-LAEs, using 
a set of 29 candidates with either spectroscopic confirmations 
\citep{fin11, cho13, son13} or a very high ($\geq$ 0.9) probability of association.
We then used the Kolmogorov-Smirnov statistic to test whether the sample's
distributions of stellar mass, age, and reddening were 
in any way different from those formed from the remaining 34 LAEs.  
Figure \ref{fig:samples} compares the empirical cumulative distribution 
functions (ECDFs) of the samples for all three SED parameters.
In all cases, the two distributions are statistically indistinguishable.
Misidentifications are therefore not biasing the results of our SED fitting.

\subsection{Size Measurements}
\label{sec:size}

To complement our SED-based estimates of stellar mass, extinction, and age, 
we also measured the sizes of the LAEs' star-forming regions, via rest-frame
UV measurements from {\sl HST\null}.  In the COSMOS field, deep F814W images 
are available, which, for the LAEs targeted in this study, sample the
rest-frame wavelength region between $\sim 2100$~\AA\ and $\sim 2800$~\AA, 
depending on redshift.  For GOODS-N LAEs, we have access to
F606W observations, which are sensitive to rest-frame wavelengths from 
$\sim 1600$~\AA\ to $\sim 2100$~\AA\null.  In both cases, we followed the 
exact same procedures described in Section 3.2 of \cite{bon09}.  After 
creating cutouts around each galaxy, we performed object identifications
and background subtraction using the routines found in SExtractor 
\citep{ber96}.  We then obtained a measure of size by using the 
IRAF\footnotemark \footnotetext{IRAF is distributed by the National Optical 
Astronomy Observatories, which are operated by the Association of Universities 
for Research in Astronomy, Inc., under cooperative agreement with
the National Science Foundation.} program {\tt phot} to determine each 
object's flux-weight centroid and magnitude through a series of circular 
apertures.  These aperture magnitudes were then used to define each LAE's half-light radius. 
As described by \citet{bon12}, the uncertainty in this type of measurement is given by
\begin{equation}
\frac{\sigma_r}{r} = 0.54 \, \frac{\sigma_f}{f}
\end{equation}
where $r$ is the half-light radius, $f$ and $\sigma_f$ are the flux and 
associated flux uncertainty, and $\sigma_r$ is the resultant error on the 
half-light radius. For our sample of LAEs as a whole, the fractional median uncertainty for the measured
half-light radius is 4\%.  We note that this measure of size is much less
sensitive to the $(1+z)^4$ effects of cosmological surface brightness dimming than
estimates based on limiting isophotes. Furthermore, the observations differ in depth (COSMOS is a single orbit while GOODS is 2.5 orbits), 
and so any biases from surface brightness limits can be found by comparing the
 half-light radii from both surveys. 
A two-sample KS-test showed that the distributions of half-light radii derived from GOODS and 
 COSMOS are not significantly different and thus we should not be concerned about effects from 
 cosmological surface brightness dimming.
 
\section{Results}
\label{sec:results}
Table~\ref{table:sed} gives the best-fit solutions to our SED fits, their 
reduced $\chi^2$ values, and the LAE's half-light radii as measured on the 
{\sl HST\/} frames.  We discuss these results below.

\subsection{Size}
\label{sec:results_size}
\begin{figure}[t]
\plottwo{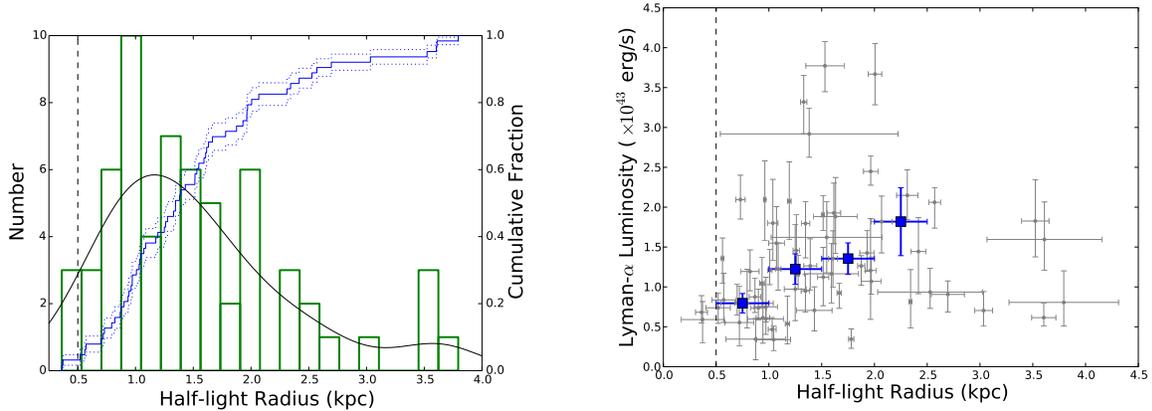}{{fig3b}.pdf}
\caption{{\it Left:} A histogram showing the distribution of
half-light radii for the HPS LAEs.  The cumulative distribution is shown 
in blue, while the kernel density estimation (KDE) of the distribution 
is in black.  The dashed line illustrates the resolution limit of the data.  
The median half-light radius of 1.35~kpc is consistent with that found for 
narrow-band selected LAEs at $z \sim 2.1$, but larger than the median at 
$z \sim 3.1$ \citep{bon12}.  {\it Right:} A comparison of half-light radius to 
Ly$\alpha$ luminosity.  The grey points are the individual measurements, while 
the dark blue squares show the median value of each bin.  The error bars in $x$ 
delineate the size of each bin, while the error bars in $y$ report the bin's 
standard error.  Although the scatter is large, the correlation for LAEs with 
$r < 2.5$~kpc is significant at the $3.4 \, \sigma$ (99.9\%) confidence
level, and for the entire sample, the trend is confirmed with 
$2.5 \, \sigma$ (99\%) confidence.
}
\label{fig:size}
\end{figure}

\begin{figure}[t]
\centering
\includegraphics[scale=0.6]{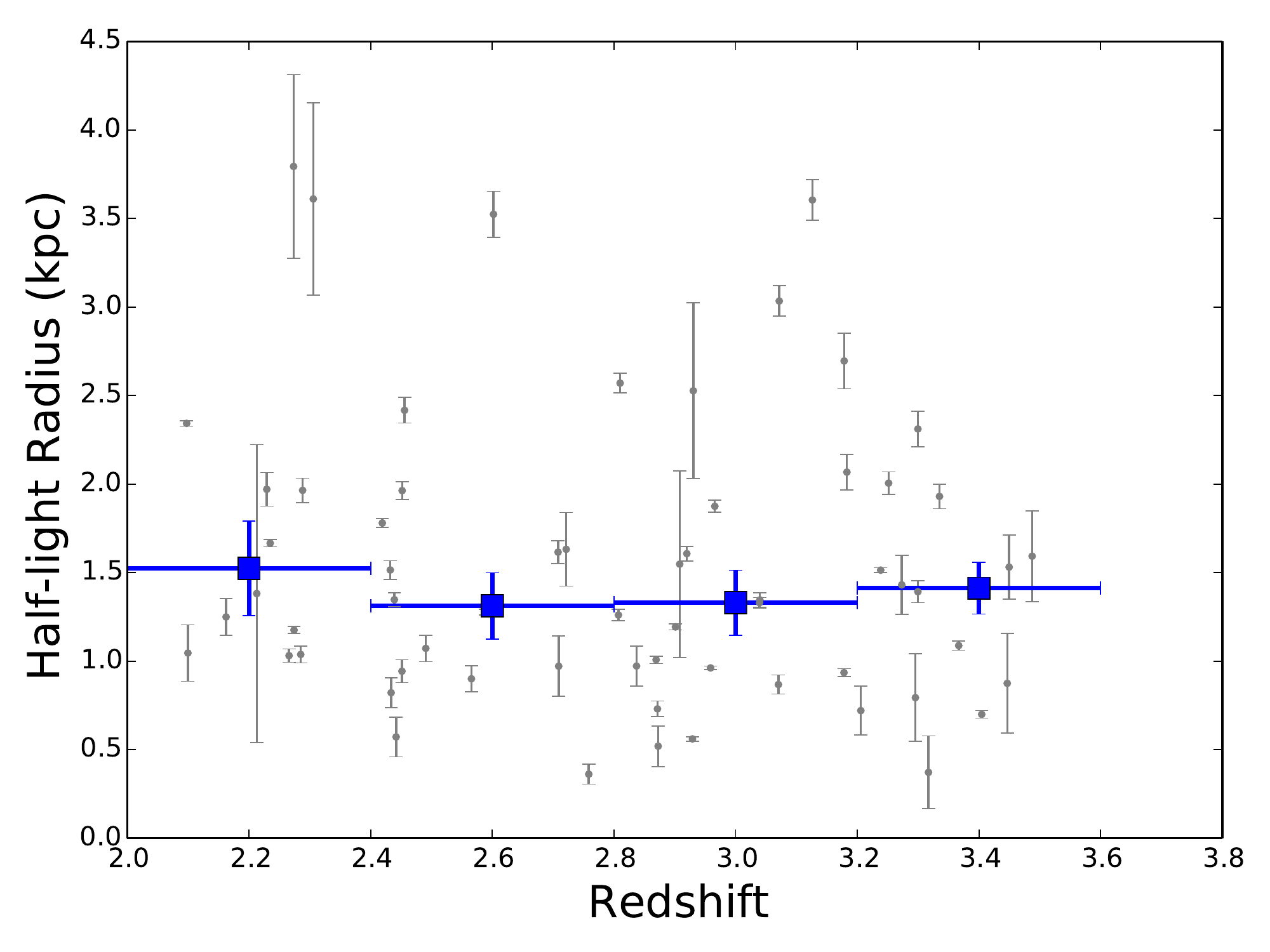}
\caption{The lack of LAE size evolution with redshift. The grey points are individual measurements
while the blue squares are binned medians. A bias in size measurement from cosmic surface brightness
dimming would manifest as a decrease in half-light radius as $(1 + z)^{4}$.  There is no evidence
for this effect.}
\label{fig:size-evol}
\end{figure}

The left-hand panel of Figure~\ref{fig:size} shows the distribution of 
half-light radii for the 63 luminous Ly$\alpha$ emitters in our 
sample.\footnotemark\footnotetext{This type of plot will be used throughout 
this work and shows a histogram, a kernel density estimation (KDE), and
 the empirical cumulative distribution function (ECDF).  All KDEs 
contained herein use a Gaussian kernel and were calculated using the 
\textit{density} function in R \citep{R}. The bandwidth for each KDE was 
found using the rule of thumb from \cite{sco92}. A simple change in the 
choice of bins for a histogram can change the interpretation of the science; 
since KDEs do not require binning, they do not suffer from this effect.  
Every ECDF plotted will also have dotted upper and lower limits, which 
represent the $1\sigma$ asymptotic errors \citep{don52}.}    From the figure, 
it is clear that the highly luminous LAEs of the HPS have a wide range of 
sizes:  while the peak of the distribution is close to 1.2 kpc, there is a
distinct tail that extends all the way out to $\sim 4$~kpc.  The median of the
distribution is $1.35^{+0.08}_{-0.10}$~kpc, where the 68\% confidence 
interval comes from a bootstrap analysis \citep{efr87}.   For comparison, the 
typical size of the narrow-band selected $2 < z < 3$ LAEs studied by 
\citet{bon12} is $\sim 1.0$~kpc, while that for $z \sim 3$~Lyman-break 
galaxies is close to 4~kpc \citep{pen10}.   As demonstrated by Spearman
rank correlation coefficient ($\rho = -0.15$) and Figure \ref{fig:size-evol}, 
these sizes show no significant correlation with redshift; this result is consistent with 
that of \citet{mal12}, who also saw no size evolution in samples of narrow-band 
selected LAEs between $2.5 < z < 6$.  This is in contrast with the strong evolution seen in the 
sizes of LBGs over the same redshift range \cite[e.g.][]{fer04}.

On the other hand, as the right-hand panel of Figure~\ref{fig:size} shows, 
there is a relation between LAE half-light radius, as measured in the 
rest-frame UV, and the amount of luminosity emitted in the Ly$\alpha$ line. 
Although the scatter in the diagram is substantial, the Spearman rank 
order coefficient reveals a positive correlation between the two 
variables ($\rho = +0.31$), which is significant at the $2.5 \, \sigma$ (99\%) 
level.  Moreover, up to a half-light radius of 2.5~kpc, $\rho=+0.45$, implying
a $3.4 \, \sigma$ (99.9\% confidence) result.   Within this range, Ly$\alpha$ 
luminosity appears to grow linearly with galaxy size with a slope of 
$0.6 \pm 0.2 \times 10^{43}$ erg s$^{-1}$ kpc$^{-1}$.  Unfortunately, the data 
outside this range are too sparsely populated to draw any conclusions.

\subsection{Stellar Mass}
\label{sec:results_mass}
\begin{figure}[t]
\plottwo{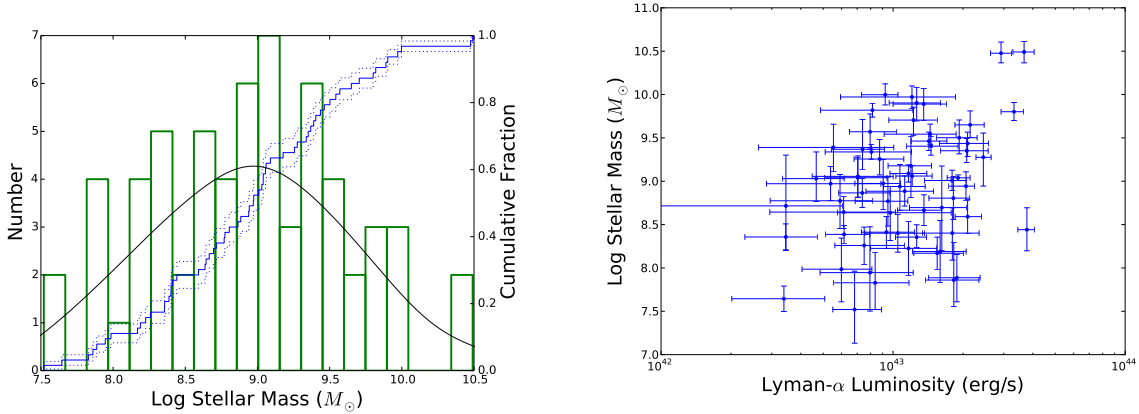}{{fig5b}.pdf}
\caption{{\it Left:}  The distribution of masses for the HPS LAEs under 
the assumption of a constant star formation rate throughout history.  The 
items plotted on the left figure are described in the caption of 
Figure~\ref{fig:size}.  The distribution is very nearly flat in the log over 
3 orders of magnitude.  {\it Right:}  a comparison of galaxy mass with 
Ly$\alpha$ luminosity.  There is no significant correlation between the 
two parameters, implying that the sample's Ly$\alpha$
flux-limit does not propagate strongly into a constraint on stellar mass.}
\label{fig:mass}
\end{figure}

At $2 < z < 3$, SED stacking analyses have produced estimates for the median LAE mass
in the range of $\sim10^8$ to $10^{10} M_{\odot}$, depending on whether the samples under
study were detectable on Spitzer/IRAC images \citep{lai08, gua11, acq11}.  In the left-hand
panel of Figure~\ref{fig:mass}, we show the distribution of {\it individual\/} LAE masses derived
from our constant star formation rate model.  Although the median mass of $\log (M/M_\odot) =
8.97^{+0.06}_{-0.17}$ lies in the range inferred from the previous stacking analyses, the data
span a factor of a thousand, from $M \sim 10^{7.5}$ to $10^{10.5} \, M_\odot$. This
distribution is consistent with that recently found by \citet{mcl14} for a set of extremely luminous
($L({\rm Ly}\alpha) > 10^{43}$~ergs~s$^{-1}$) $z \sim 3.1$ LAEs found via narrow-band imaging.  
Interestingly if we fit our LAE mass distribution with a power law over the range 
$8.5 < \log M/M_{\odot} < 10.5$, then the most-likely slope, $\alpha = -1.3 \pm 0.1$, is similar, or 
perhaps only slightly shallower than the slope usually adopted for the epoch's galaxy
mass function \citep[\eg][]{muzzin+13, tomczak+13}.  This suggests that bright Ly$\alpha$
emitters are drawn from an underlying distribution that is not strongly dependent on
stellar mass.   Moreover, as the right-hand panel of Figure~\ref{fig:mass} demonstrates,
there is no obvious correlation between stellar mass and Ly$\alpha$ luminosity:
at any $\log L$, one can find LAEs of all masses, and galaxies of any given
mass can have a wide range of Ly$\alpha$ luminosity.   Unless this
behavior abruptly changes at low Ly$\alpha$ luminosity, it would appear that
large-volume LAE surveys are an excellent way of sampling virtually the entire
range of the high-redshift galaxy mass function.

One additional feature of Figure~\ref{fig:mass} worth noting is the absence of 
LAEs with masses less than $\sim 5 \times 10^7 M_{\odot}$.   There are two
possible reasons for this.  The first is the depth of the imaging:  there are
seven sources in the HPS survey for which \citet{ada11} could find no obvious 
counterpart in the rest-frame UV\null.  An examination of the CANDELS frames 
and grism spectra from the 3D-HST program \citep{van13} reveals that only one 
of these missing objects has any detectable flux in the rest-frame optical.
(The counterpart of HPS~266 is detected at  $\alpha(2000) = $10:00:29.818,
$\delta(2000) = $+2:18:49.20.)  The missing 6 objects may therefore be part of the 
extreme low-mass tail of the mass function.

A second possible explanation for the missing lower-mass galaxies comes from 
the limitations imposed by our input physics.  Formally, the population 
synthesis models of \citet{bru03} are applicable to stellar systems with ages 
between $10^5 < t < 2 \times 10^{10}$~yr.
However, the CB07 (and BC03) models used in our analysis, and the 
Padova isochrones upon which they are based, were developed using stars with ages of $10^{6.6}$ 
years and older \citep{con13}.  Any system younger than this must therefore 
be subject to  greater systematic uncertainties.  The lack of systems with 
masses below $\sim 5 \times 10^7 M_{\odot}$ may therefore be an artifact of 
our SED fitting.  Furthermore, since our SED fits assume a constant
star formation rate, very low mass systems almost certainly have very 
young ages, further increasing the potential for systematic errors at the
very faint end of our sample.

\begin{figure}[t]
\plottwo{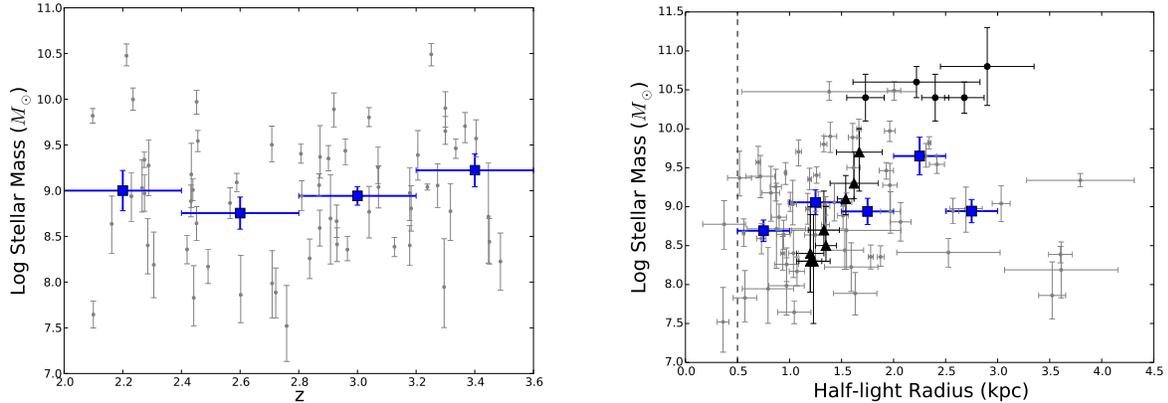}{{fig6b}.pdf}
\caption{The left panel displays stellar masses versus redshift, with the 
individual LAEs shown in grey and the median values for each redshift bin 
plotted as blue squares. The right panel
similarly compares stellar mass to half-light radius.  The dashed line 
illustrates the limit of our spatial resolution.  In both figures, the 
error bars in $x$ illustrate the sizes of the bins, while the errors in 
$y$ report the standard error within each bin. The black circles are 
averages of various continuum selected galaxies from \cite{mos11} 
and the black triangles are results from stacks by \cite{bon12}.}
\label{fig:mass-zr}
\end{figure}

Figure~\ref{fig:mass-zr} plots our derived stellar masses against two 
parameters, LAE size and redshift.  Neither shows a significant trend.
While \citet{bon12} did find a correlation between mass and half-light radius, their
analysis dealt with stacked images, not individual SEDs.  Our null result also agrees with
that found for surveys of UV-bright galaxies in the same redshift range.  LAE
sizes (as measured in the UV continuum) and stellar masses do not seem to be related 
\citep[e.g.,][]{mos11}.

\subsection{$E(B-V)$}
\label{sec:ebv}
\begin{figure}[t]
\plottwo{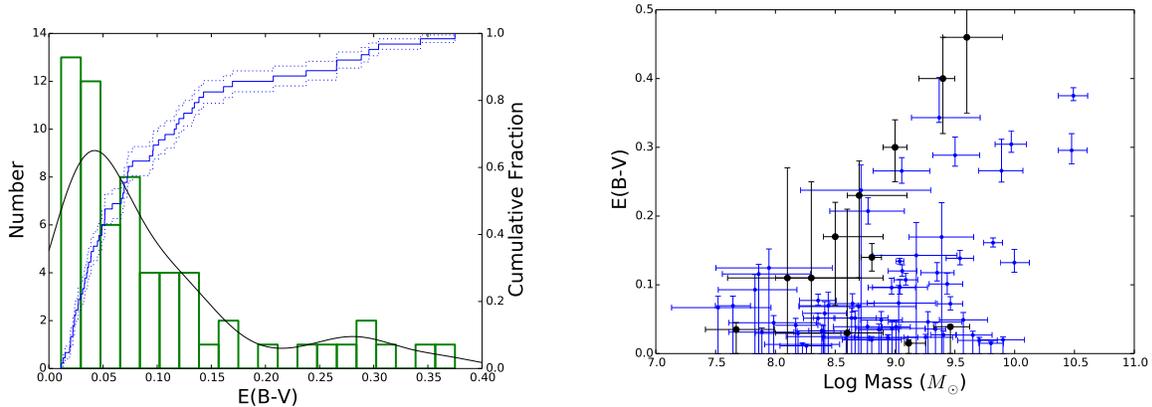}{{fig7b}.pdf}
\caption{{\it Left:  }The distribution of stellar reddenings derived from our 
constant SFR models. The items plotted are described in the caption of
Figure~\ref{fig:size}.  The median of the distribution is $E(B-V) = 0.07$, but 
there is a tail that extends out to $E(B-V) \sim 0.4$.  {\it Right:  }  The 
stellar reddening of our LAEs as a function of galaxy mass. The black circles 
represent the results of stacking analyses for LAEs at $z \sim 2.1$ and $z \sim 3.1$
\citep{acq11, acq12b, gua11}.  Low mass objects 
are uniformly dust-poor, but objects with $M > 10^{9} \, M_{\odot}$ can have 
a wide range of internal extinction. Stellar mass and extinction are correlated 
with a Spearman rank coefficient of $\rho = 0.3$, indicating $2.4 \, \sigma$ (98.5\%) significance.}
\label{fig:ebv}
\end{figure}

Since Ly$\alpha$ photons resonantly scatter off interstellar hydrogen, even a 
small amount of extinction can reduce the emergent emission-line flux by several 
orders of magnitude.  Thus, it has long been argued that LAEs are mostly 
young, metal-poor objects with very little dust \citep[\eg][]{gaw06, mao07}. 
Nevertheless, evidence for the existence of dust within LAEs has been seen in the work of \citet{fin09}
among others.  As the left-hand panel of Figure~\ref{fig:ebv} illustrates,
our data demonstrate that, indeed, most LAEs are dust-poor.  
Based on our SED analyses, half of the HPS LAEs have internal stellar 
reddenings $E(B-V) < 0.07$, though there is a tail that
extends all the way out to $E(B-V) \sim 0.4$.  The median of the $E(B-V)$ 
distribution is $0.067^{+ 0.003}_{-0.018}$.  Notably, all the high-extinction
objects are drawn from the high-mass end of the LAE mass function:  
every LAE with $E(B-V) > 0.25$ has a mass greater than $\sim 10^9 M_{\odot}$.  
This agrees with correlations between mass and extinction seen in both the
local universe and at high redshift \citep[e.g.,][]{gar10,kas13}.

\begin{figure}[t]
\plottwo{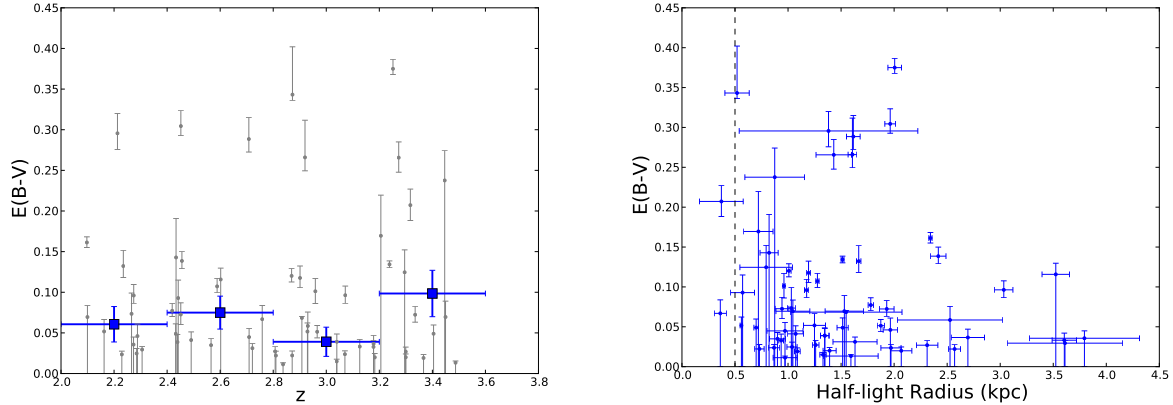}{{fig8b}.pdf}
\caption{Comparisons of stellar reddening with redshift (left) and 
half-light radius (right).  The grey points show individual LAEs while the 
blue points illustrate the median values within each bin.  The dashed line 
shows the limit of our spatial resolution.  Neither diagram displays 
a significant trend.}
\label{fig:dust_zr}
\end{figure}

Figure~\ref{fig:dust_zr} displays our estimates of differential reddening 
versus redshift and galaxy size.  There are no significant trends in either 
diagram. Unlike \citet{gua11} and \citet{acq11}, we see no evidence for any 
change in the mean reddening of the LAE population versus redshift, and, 
unlike \citet{bon12}, we see no correlation between $E(B-V)$ and half-light
radius.  Again, we caution that the LAE samples considered here are 
more luminous than those derived from narrow-band surveys, and by studying
individual, rather than stacked spectra, we are avoiding many of the systematic 
difficulties that complicate the interpretation of previous measurements 
\citep{var13}.

\subsection{Age}
\label{sec:age}
\begin{figure}[t]
\plottwo{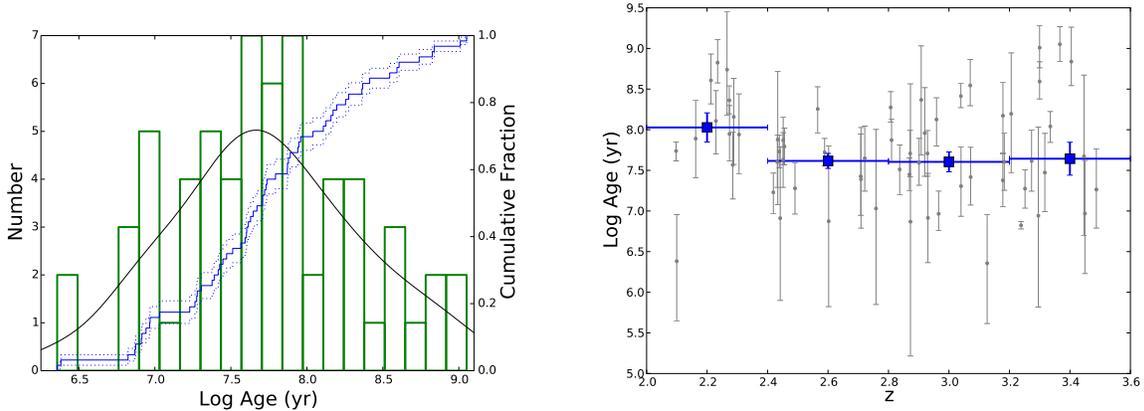}{{fig9b}.pdf}
\caption{\textit{Left:} The age distribution of the HPS LAEs. The ages extend over $\sim 2.5$ dex. 
\textit{Right:} The distribution of LAE ages versus redshift, with individual objects plotted in grey and 
binned medians in blue.  There is no evidence for evolution in the LAE population.}
\label{fig:age}
\end{figure}

Figure~\ref{fig:age} displays the age distribution for the HPS LAEs,
under the simplest assumption of a constant star-formation rate history.   The two figures
together support the stacking analyses of \cite{lai08}, \cite{gua11}, and \cite{acq11}, which
argued that LAEs are relatively young, with ages between $\sim 10^7$ and $10^9$ years.  The median
age of the HPS sample is $\log t = 7.96^{+0.19}_{-0.14}$, and
just $\sim 3\%$ have ages greater than a Gyr.   Interestingly, there no evidence for 
evolution in the sample. This disagrees with result of \citet{gua11}, who found narrow-band selected
$z = 2.06$ LAEs to be older and dustier than their $z = 3.1$ counterparts.  It is also
in conflict with the re-analysis by \citet{acq12b}, who concluded that these same LAEs were
``growing younger'' with time.  A likely explanation for this discrepancy lies in the details of the stacking procedure used by both groups, as slight differences can produce discrepant results \citep[see][]{var13}.  
Our analysis of individual LAEs avoids that pitfall.

\subsection{Star Formation Rates and the Main Sequence of Galaxies} 
\label{sec:sfr_history}
The observed star formation rate of an LAE can be derived several ways: from 
the luminosity of its Ly$\alpha$ emission line via the assumption of Case~B 
recombination, 
\begin{equation}
{\rm SFR(Ly}\alpha) = 9.1 \times 10^{-43} \, L({\rm Ly}\alpha) \
M_{\odot}~{\rm yr}^{-1}
\label{eq:ly-alpha}
\end{equation}
\citep{bro71, ken98}, from the extinction-corrected flux density of the UV continuum 
between 1500~\AA\ and 2800~\AA
\begin{equation}
{\rm SFR(UV)} = 1.4 \times 10^{-28} \, L_{\nu} \ M_{\odot}~{\rm yr}^{-1}
\label{eq:uv_sfr}
\end{equation}
\citep{ken98}, and via the fit to the object's spectral energy distribution 
(a value which is largely dependent on the UV emission, but which  
may also include factors such as age).  

\begin{figure}[t]
\plotone{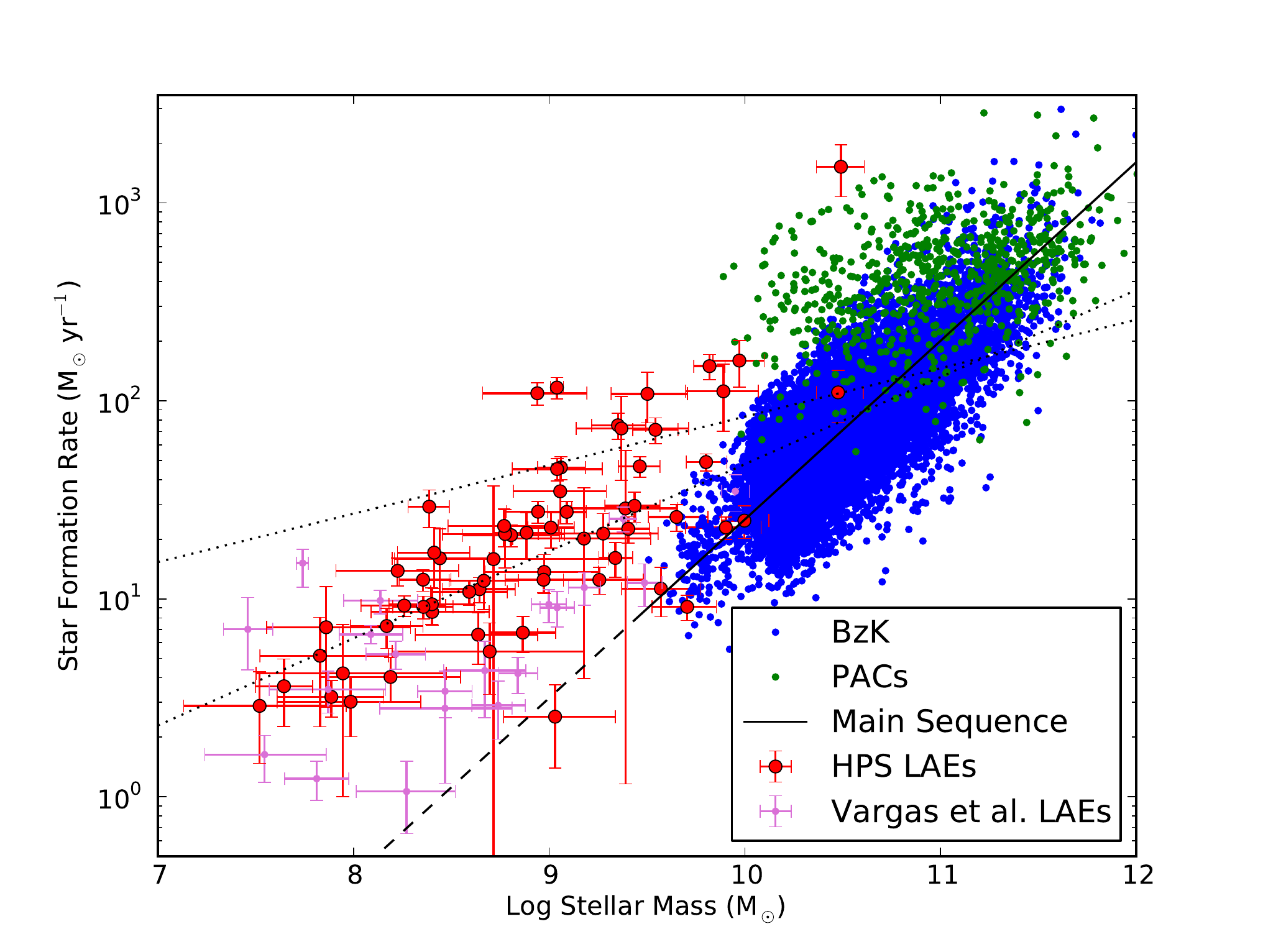}
\caption{Dust corrected UV star formation rate plotted against stellar mass
for various high-redshift galaxies.  Our luminous, spectroscopically-selected
LAEs are shown in red, and narrow-band selected LAEs from \citet{var13} are
in pink.  The blue and green points show higher-mass galaxies from the 
\citet{rod11} survey of COSMOS and GOODS-N:  blue are continuum-selected 
($BzK$) galaxies, while green shows sub-mm bright systems found by the {\sl 
Herschel-}PACS instrument.  The solid black line is the star-forming ``main 
sequence'' defined by \citet{dad07} for $1.5 < z < 2.5$; the extrapolation of 
this line to lower SFRs is shown with a dashed line.  LAEs and sub-mm
galaxies lie above this relation, along the main sequences found by 
\cite{whi12} for $z=2.0$ (lower dotted line) and $z=3.5$ (upper dotted line).}
\label{fig:mainseq}
\end{figure}

Figure \ref{fig:mainseq} plots our dust-corrected UV SFRs against stellar
mass, and compares these data to those obtained from other samples of 
high-$z$ galaxies.   From the figure, it is clear that
each selection technique identifies galaxies in a different region of 
the mass-SFR diagram.  LAEs are primary low-mass, low-SFR objects
that lie above the star-forming main sequence found by \citet{dad07}, 
in a region of the diagram consistent with measurements of galaxy main-sequence 
evolution \citep{whi12}.  Both IFU and narrow-band selected LAEs fall in 
this same region, confirming that both discovery techniques trace the 
same population.  UV continuum-detected ($BzK$) galaxies are higher-mass, 
high-SFR objects, while {\sl Herschel}-PACS-selected sub-mm galaxies are 
 high-mass objects that, like LAEs, fall predominantly above the 
star-forming main sequence \citep{rod11}.  As the distributions of LAEs 
and sub-mm galaxies abut each other, it is tempting to associate the two 
classes. If sub-mm systems are the results of merger-driven starbursts \citep{con03}, 
then LAEs could potentially be their low mass and low-dust counterparts:  in general, 
a starburst event will move a galaxy up and slightly to the right on this diagram. \citet{gro11}, 
however, see no strong evidence for interactions in LAE morphologies.

\subsection{Ly$\alpha$ Escape Fractions and the q-factor}
\label{sec:escape}
Like the UV flux density, the H$\alpha$ emission line is a well-known and 
well-understood star formation rate indicator \citep{ken98}.  Since under 
Case~B recombination, roughly three quarters of all Balmer transitions 
produce a Ly$\alpha$ photon, this means that Ly$\alpha$ should also 
be a robust tracer of star formation.  Thus, {\it if\/} all the Ly$\alpha$ 
and UV continuum photons escape into intergalactic space, Ly$\alpha$ and the 
dust-corrected UV continuum should be well-correlated.  Systematic 
deviations from a one-to-one relation then measure the escape fraction
of Ly$\alpha$ photons.  Note that this differential procedure sidesteps 
the issue of whether the reddening derived from the stars is the same as 
that for the gas, but it does assume that both the UV and Ly$\alpha$
emission is isotropic.  It is also susceptible to a systematic error 
associated with the timescale of star formation.  The nominal 
conversion between UV emission and star formation rate 
(equation~\ref{eq:uv_sfr}) assumes a timescale for star formation that is
ten times longer than that for emission line gas.  If a system is undergoing
a rapid burst of star formation ($\tau < 10^8$~yr), its two 
SFRs indicators may not be comparable.
Nevertheless, the ratio of UV flux to Ly$\alpha$ can provide 
constraints on the radiative transfer of the emission line.

The question of the dependence of Ly$\alpha$ escape fraction on SFR and redshift
has been recently discussed in \cite{dij13}.  This paper presents two models: 
one in which the escape fraction is independent of star formation rate, and 
a second where the escape fraction decreases as the SFR increases. 
We find a significant inverse correlation between stellar mass and escape 
fraction, $\rho = -0.54$ ($4.5 \, \sigma$ or $> 99.999\%$ significance) which 
supports the second model; this is shown in Figure \ref{fig:escape}.  
Unsurprisingly, the Ly$\alpha$ escape fraction also correlates with differential extinction, 
as mass and $E(B-V)$ are coupled (see~Fig.~\ref{fig:ebv}).  We note
that the median escape fraction of our sample, $\sim 0.5$ (or 0.6, if we 
use the SED-based SFRs), is somewhat larger than the 0.29 value found by
\citet{bla11} using the same sample of LAEs.  Most of this difference is
due to the use of deeper CANDELS data, which greatly improves the 
photometry and fixes the slope of the rest-frame UV.

\begin{figure}[t]
\plottwo{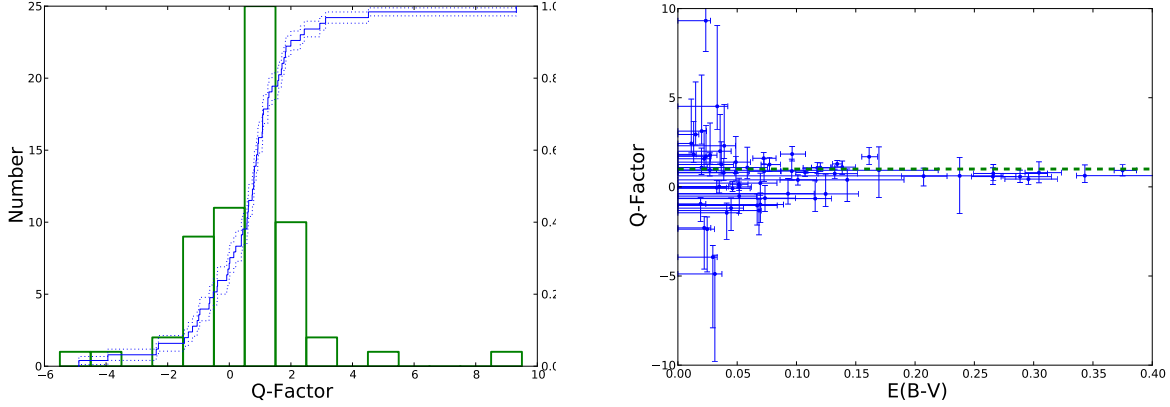}{{fig11b}.pdf}
\caption{\textit{Left:} A histogram of the $q$-parameters for our
spectroscopically selected LAEs. \textit{Right:}  These same $q$ values as 
a function of $E(B-V)$.  The data are biased, in that systems with large
values of $q$ will have Ly$\alpha$ suppressed below the limit of 
detectability.  All dusty systems have $q$ near 1, while systems with
small $E(B-V)$ can have a range of $q$ values.}
\label{fig:q}
\end{figure}

\begin{figure}[t]
\plottwo{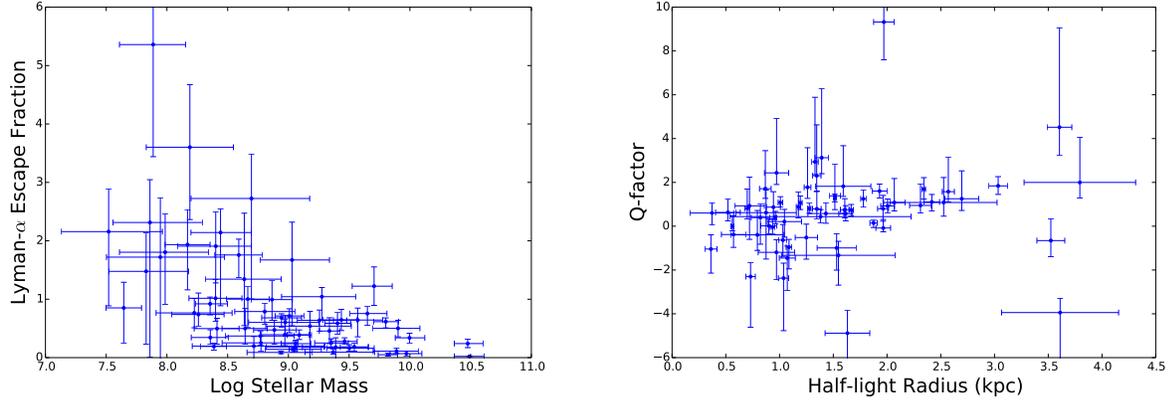}{{fig12b}.pdf}
\caption{\textit{Left:} The escape fraction of Ly$\alpha$ photons as a function of stellar mass.  Note
the negative correlation.  Values greater than one can be due to an anisotropic ISM, as Ly$\alpha$ may be
escaping through voids blown out by supernovae or other feedback.   The decline in the maximum escape
fraction with stellar mass supports this model, as it would take much more energy to blow out a hole in a 
massive galaxy. \textit{Right:} The correlation between half-light radius and $q$; galaxies with small or 
negative $q$ factors are physically smaller.  This is again consistent with the idea with anisotropic
Ly$\alpha$ emission, as it would be easier for supernovae to create holes in smaller, less massive
galaxies.}
\label{fig:escape}
\end{figure}

Perhaps a more useful way of looking at the radiative transfer problem is
through the variable $q$, which relates Ly$\alpha$ optical depth to that 
of the stellar continuum at 1216~\AA\null.  As 
defined by \citet{fin08},
\begin{equation}
q = \frac{\tau_{Ly\alpha}}{\tau_{1216}}
\end{equation}
where $\tau_{\lambda} = 0.921 \, k_{\lambda} E(B-V)$ and $k_{1216} = 5.27$
under the empirical reddening law described by \citet{cal01}.  Figure~\ref{fig:q} shows the
distribution of $q$ values and the behavior of $q$ with $E(B-V)$.  
Interestingly, at large reddenings $q$ never gets much above 1, suggesting
that in these systems, the Ly$\alpha$ emitting gas and the UV starlight
are seeing the same amount of extinction.  We do expect that galaxies with
large Ly$\alpha$ optical depths will be censored out of our LAE
sample.  However, in dust-rich systems it appears that, if Ly$\alpha$
escapes, it does so with very few resonant scatterings. This is consistent with
models that involve strong winds, such as that proposed by \cite{ver08}.
On the other hand, at low reddenings, we see a large range of $q$ values.  Systems
with $q < 0$ imply anisotropic emission, a top-heavy IMF, or
a very young starburst, where the UV luminosity to SFR conversion breaks 
down.   As expected, we find that the half-light radius and $q$-factor are 
positively correlated, with a Spearman rank order coefficient of $\rho = 0.35$
(99.5\% confidence).   This correlation is shown in the right hand panel of Figure \ref{fig:escape}. 
A small size could lead to less homogeneity and thus more 
opportunities for Ly$\alpha$ to undergo anisotropic radiative transfer. 

\section{Conclusion}
\label{sec:conclusion}

Using broadband photometric data which extends from the rest-frame UV through
to the near-IR, we have been able to measure the stellar masses, reddenings, 
and sizes for a sample of 63 luminous LAEs found in the HETDEX 
Pilot Survey.  Our fits demonstrate that, contrary to popular belief, 
Ly$\alpha$ emitters are not exclusively low mass objects.  In fact,  
HPS-selected LAEs are quite heterogeneous, and are drawn from almost the 
entire stellar mass range of high-redshift galaxies.  Moreover, there is a striking 
similarity between the mass function of LAEs and the mass function 
expected for the galactic star-forming population as a whole.  This fact,
and the lack of correlation between Ly$\alpha$ luminosity and stellar mass,
suggests that searches for Ly$\alpha$ emission are excellent way of sampling 
a large fraction of the mass function of high-redshift star-forming galaxies.

Ly$\alpha$-emitting galaxies occupy a different part of stellar mass-SFR parameter space
than that of galaxies found by other methods. Like the higher-mass sub-mm galaxies, LAEs fall
above the main sequence of star-forming galaxies found by \cite{dad07}. This suggests that there
is a different slope for the main sequence of star-bursting galaxies. Interestingly, LAEs do fall along the
main sequence defined by \cite{whi12}, though the $\sim 2$ dex extrapolation required to reach their 
masses introduces significant uncertainty.
Due to the various selection effects at work, the connection between
the various classes of star-forming galaxies is murky at best.

We also find that the range in observed $q$-factors is dependent on the 
reddening, with the widest range of $q$-values occurring at low extinction.  Interestingly,
the observed values of $q$ tend to unity as the reddening (or mass) increases, 
suggesting that in these objects, Ly$\alpha$ photons are not undergoing a large number 
of scattering events.  This
strongly implies that winds are an important component in the making of high-mass
LAEs.  Furthermore, we find that the half-light radius and the $q$-factor
are positively correlated, implying that Ly$\alpha$ emission is enhanced in very small 
objects.

\acknowledgments

\section*{Acknowledgments}

We thank the referee for their helpful comments.
We also thank Joshua Adams for the use of his photometry from the HETDEX Pilot Survey. 
We acknowledge the Research Computer and Cyberinfrastructure Unit of Information 
Technology Services,  and in particular W. Brouwer at The Pennsylvania State 
University for providing computational
support and resources for this project.  
This work is based on observations taken by the
CANDELS Multi-Cycle Treasury Program with the NASA/ESA HST, which is 
operated by the 
Association of Universities for Research in Astronomy, Inc., under NASA 
contract NAS5-26555.  
The work was also partially supported by NSF grants AST 09-26641 and 
AST 10-55919. The Institute for Gravitation and the
Cosmos is supported by the Eberly College of Science and the Office 
of the Senior Vice President 
for Research at the Pennsylvania State University.

This research has made use of NASA's Astrophysics Data System and the 
python packages \texttt{IPython}, \texttt{NumPy}, \texttt{SciPy}, and 
\texttt{matplotlib}. 

{\it Facilities:} \facility{CFHT}, \facility{HST}, \facility{Mayall} \facility{Smith}, \facility{Spitzer (IRAC)}, \facility{Subaru}, \facility{UH:2.2m}, \facility{UKIRT}

\clearpage

\begin{deluxetable}{cccrccc}
\tabletypesize{\scriptsize}
\tablewidth{0pt}
\tablecaption{COSMOS Field Photometry \label{table:cosmosphot}}
\tablehead{
\colhead{Telescope} & \colhead{Instrument} & \colhead{Filter} & \colhead{Central $\lambda$} & \colhead{Original} & \colhead{Photometry} & \colhead{$5 \sigma$ Limits} \\ 
\colhead{} & \colhead{} & \colhead{} & \colhead{(\AA)} & \colhead{Survey} & \colhead{} & \colhead{(AB)} }
\startdata
CFHT & Megaprime & u* & 4065 & COSMOS & \cite{ada11} & 26.5 \\
Subaru & Suprime-Cam & B & 4788 &  COSMOS &  \cite{ada11} & 27.4\\
Subaru & Suprime-Cam & V & 5730 &  COSMOS &\cite{ada11} & 27.2 \\
Subaru & Suprime-Cam & r+ & 6600 &  COSMOS &\cite{ada11} & 26.9 \\
HST & ACS & F814W & 7461 & CANDELS & \cite{son13} & 27.5 \\
Subaru & Suprime-Cam & i+ & 7850 &  COSMOS &\cite{ada11} & 26.9 \\
Subaru & Suprime-Cam & z+ & 8700 &  COSMOS &\cite{ada11} & 25.6 \\
UKIRT & WFCAM & J & 12850 &  COSMOS &\cite{ada11} & 23.6 \\
HST & WFC3 & F125W & 13250 & CANDELS & \cite{son13} & 26.4 \\
HST & WFC3 & F160W & 14460 & CANDELS & \cite{son13} & 26.5 \\
CFHT & WIRCAM & K & 21400 &  COSMOS &\cite{ada11} & 23.6 \\
Spitzer & IRAC & Channel 1 & 37440 & S-COSMOS & This paper & 23.9\\
Spitzer & IRAC & Channel 2 & 44510 & S-COSMOS & This paper & 23.3\\
Spitzer & IRAC & Channel 3 & 59950 & S-COSMOS & This paper & 21.3\\
Spitzer & IRAC & Channel 4 & 84870 & S-COSMOS & This paper & 21.0\\
\enddata
\tablecomments{CANDELS covers 32 of 42 objects in this field}
\end{deluxetable}

\begin{deluxetable}{cccrccc}
\tabletypesize{\scriptsize}
\tablewidth{0pt}
\tablecaption{GOODS-N Field Photometry \label{table:goodsnphot}}
\tablehead{
\colhead{Telescope} & \colhead{Instrument} & \colhead{Filter} & \colhead{Central $\lambda$} & \colhead{Original} & \colhead{Photometry} & \colhead{$5 \sigma$ Limits} \\ 
\colhead{} & \colhead{} & \colhead{} & \colhead{(\AA)} & \colhead{Survey} & \colhead{} & \colhead{(AB)} }
\startdata
Mayall & MOSAIC & U & 4065 & GOODS & \cite{ada11} & 27.1 \\
HST & ACS & F435W & 4570 & CANDELS & \cite{fin13} & 27.8 \\
Subaru & Suprime-Cam & B & 4788 &  GOODS &  \cite{ada11} & 26.9 \\
Subaru & Suprime-Cam & V & 5730 &  GOODS &\cite{ada11} & 26.8 \\
Subaru & Suprime-Cam & r+ & 6600 &  GOODS &\cite{ada11} & 26.6 \\
HST & ACS & F606W & 6690 & CANDELS & \cite{fin13} & 27.6 \\
HST & ACS & F775W & 7380 & CANDELS &   \cite{fin13} & 27.5 \\
Subaru & Suprime-Cam & i+ & 7850 &  GOODS &\cite{ada11} & 25.6 \\
HST & ACS & F850LP & 8610 & CANDELS &   \cite{fin13} & 27.3 \\
Subaru & Suprime-Cam & z+ & 8700 &  GOODS &\cite{ada11} & 25.4 \\
HST & WFC3 & F105W & 11783 & CANDELS &  \cite{fin13} & 26.6 \\
HST & WFC3 & F125W & 13250 & CANDELS &  \cite{fin13} & 26.4 \\
HST & WFC3 & F160W & 14460 & CANDELS &  \cite{fin13} & 26.5 \\
UH 2.2-m & QUIRC & H+K' & 20200 &  GOODS &\cite{ada11} & 22.1 \\
Spitzer & IRAC & Channel 1 & 37440 & GOODS & This paper & 23.9\\
Spitzer & IRAC & Channel 2 & 44510 & GOODS & This paper & 23.3\\
Spitzer & IRAC & Channel 3 & 59950 & GOODS & This paper & 21.3\\
Spitzer & IRAC & Channel 4 & 84870 & GOODS & This paper & 21.0\\
\enddata
\tablecomments{CANDELS covers 17 of 21 objects in this field}
\end{deluxetable}

\begin{deluxetable}{ccccccc}
\tablewidth{0pt}
\tablecaption{SED Fitting Results \label{table:sed}}
\tablehead{
\colhead{Galaxy ID} & \colhead{Redshift} & \colhead{Log Stellar Mass} & \colhead{Log Age} &
&\colhead{Half-light Radius}  \\ \colhead{ } & \colhead{ }
&\colhead{($M_\odot$)} &\colhead{(yr)} &\colhead{$E(B-V)$} &\colhead{(kpc)} 
&\colhead{$\chi^2_{\nu}$} }
\startdata
HPS150 & $2.90$ & $9.35_{-0.13}^{+0.14}$ & $7.59_{-0.28}^{+0.30}$ & $0.118_{-0.012}^{+0.015}$ & $1.19\pm0.02$ & $4.1$ \\
HPS153 & $2.71$ & $9.50_{-0.19}^{+0.20}$ & $7.43_{-0.48}^{+0.52}$ & $0.289_{-0.016}^{+0.026}$ & $1.62\pm0.06$ & $1.5$ \\
HPS154 & $2.87$ & $9.37_{-0.23}^{+0.34}$ & $6.87_{-1.65}^{+1.70}$ & $0.343_{-0.007}^{+0.059}$ & $0.52\pm0.11$ & $3.5$ \\
HPS161 & $3.25$ & $10.49_{-0.13}^{+0.12}$ & $7.28_{-0.24}^{+0.23}$ & $0.375_{-0.007}^{+0.012}$ & $2.01\pm0.06$ & $1.0$ \\
HPS164 & $2.45$ & $9.97_{-0.14}^{+0.13}$ & $7.87_{-0.30}^{+0.28}$ & $0.304_{-0.012}^{+0.019}$ & $1.96\pm0.05$ & $2.9$ \\
HPS168 & $3.45$ & $8.44_{-0.24}^{+0.25}$ & $6.97_{-0.74}^{+0.65}$ & $0.070_{-0.070}^{+0.020}$ & $1.53\pm0.18$ & $2.8$ \\
HPS174 & $3.45$ & $8.72_{-0.51}^{+0.59}$ & $7.67_{-0.98}^{+1.00}$ & $0.238_{-0.238}^{+0.037}$ & $0.87\pm0.28$ & $3.2$ \\
HPS182 & $2.43$ & $9.18_{-0.36}^{+0.34}$ & $7.88_{-0.80}^{+0.84}$ & $0.143_{-0.143}^{+0.048}$ & $0.82\pm0.09$ & $5.0$ \\
HPS183 & $2.16$ & $8.64_{-0.32}^{+0.30}$ & $7.89_{-0.48}^{+0.47}$ & $0.052_{-0.052}^{+0.015}$ & $1.25\pm0.10$ & $2.3$ \\
HPS184 & $3.21$ & $9.39_{-0.27}^{+0.27}$ & $8.20_{-0.72}^{+0.75}$ & $0.170_{-0.170}^{+0.050}$ & $0.72\pm0.14$ & $1.4$ \\
HPS189 & $2.45$ & $8.64_{-0.19}^{+0.18}$ & $7.63_{-0.34}^{+0.33}$ & $0.073_{-0.012}^{+0.014}$ & $0.94\pm0.06$ & $1.0$ \\
HPS194 & $2.29$ & $9.28_{-0.33}^{+0.28}$ & $8.16_{-0.59}^{+0.47}$ & $0.046_{-0.046}^{+0.015}$ & $1.96\pm0.07$ & $7.9$ \\
HPS197 & $2.44$ & $7.83_{-0.31}^{+0.35}$ & $6.91_{-1.01}^{+0.88}$ & $0.093_{-0.093}^{+0.022}$ & $0.57\pm0.11$ & $1.1$ \\
HPS205 & $2.91$ & $8.70_{-0.50}^{+0.48}$ & $8.37_{-0.68}^{+0.66}$ & $0.068_{-0.068}^{+0.002}$ & $1.55\pm0.53$ & $6.2$ \\
HPS207 & $2.71$ & $7.99_{-0.38}^{+0.36}$ & $7.39_{-0.60}^{+0.64}$ & $0.045_{-0.045}^{+0.010}$ & $0.97\pm0.17$ & $2.8$ \\
HPS210 & $3.49$ & $8.23_{-0.31}^{+0.31}$ & $7.26_{-0.47}^{+0.50}$ & $0.013_{-0.013}^{+0.002}$ & $1.59\pm0.26$ & $3.1$ \\
HPS213 & $3.30$ & $9.90_{-0.21}^{+0.18}$ & $9.01_{-0.25}^{+0.27}$ & $0.020_{-0.020}^{+0.004}$ & $1.39\pm0.06$ & $9.3$ \\
HPS214 & $3.30$ & $7.95_{-0.44}^{+0.53}$ & $6.94_{-1.12}^{+1.09}$ & $0.125_{-0.125}^{+0.028}$ & $0.79\pm0.25$ & $0.4$ \\
HPS223 & $2.31$ & $8.19_{-0.36}^{+0.36}$ & $7.94_{-0.49}^{+0.50}$ & $0.029_{-0.029}^{+0.004}$ & $3.61\pm0.54$ & $3.2$ \\
HPS229 & $3.04$ & $9.80_{-0.10}^{+0.11}$ & $8.41_{-0.14}^{+0.15}$ & $0.015_{-0.015}^{+0.003}$ & $1.33\pm0.03$ & $6.3$ \\
HPS231 & $2.72$ & $7.89_{-0.28}^{+0.27}$ & $7.65_{-0.40}^{+0.40}$ & $0.031_{-0.031}^{+0.006}$ & $1.63\pm0.21$ & $2.1$ \\
HPS244 & $2.10$ & $7.64_{-0.15}^{+0.15}$ & $6.38_{-0.74}^{+0.57}$ & $0.070_{-0.070}^{+0.014}$ & $1.05\pm0.16$ & $0.5$ \\
HPS249 & $3.27$ & $9.06_{-0.24}^{+0.24}$ & $7.62_{-0.37}^{+0.38}$ & $0.266_{-0.018}^{+0.019}$ & $1.43\pm0.17$ & $3.0$ \\
HPS251 & $2.29$ & $8.40_{-0.31}^{+0.29}$ & $7.57_{-0.42}^{+0.40}$ & $0.025_{-0.025}^{+0.006}$ & $1.04\pm0.05$ & $4.4$ \\
HPS253 & $3.18$ & $8.81_{-0.25}^{+0.25}$ & $7.60_{-0.35}^{+0.35}$ & $0.020_{-0.020}^{+0.005}$ & $2.07\pm0.10$ & $2.1$ \\
HPS256 & $2.49$ & $8.17_{-0.18}^{+0.19}$ & $7.28_{-0.32}^{+0.31}$ & $0.041_{-0.041}^{+0.010}$ & $1.07\pm0.07$ & $2.9$ \\
HPS258 & $2.81$ & $8.94_{-0.16}^{+0.17}$ & $7.87_{-0.25}^{+0.26}$ & $0.022_{-0.022}^{+0.005}$ & $2.57\pm0.06$ & $7.3$ \\
HPS263 & $2.43$ & $8.88_{-0.17}^{+0.18}$ & $7.60_{-0.31}^{+0.34}$ & $0.049_{-0.049}^{+0.012}$ & $1.51\pm0.05$ & $3.8$ \\
HPS269 & $2.57$ & $8.87_{-0.17}^{+0.17}$ & $8.25_{-0.30}^{+0.27}$ & $0.035_{-0.035}^{+0.008}$ & $0.90\pm0.07$ & $3.0$ \\
HPS274 & $2.87$ & $9.06_{-0.12}^{+0.12}$ & $7.45_{-0.20}^{+0.21}$ & $0.120_{-0.007}^{+0.009}$ & $1.01\pm0.02$ & $4.3$ \\
HPS283 & $3.30$ & $9.65_{-0.14}^{+0.16}$ & $8.59_{-0.21}^{+0.24}$ & $0.027_{-0.027}^{+0.006}$ & $2.31\pm0.10$ & $2.8$ \\
HPS286 & $2.23$ & $8.94_{-0.28}^{+0.25}$ & $8.11_{-0.41}^{+0.37}$ & $0.024_{-0.024}^{+0.004}$ & $1.97\pm0.10$ & $24.3$ \\
HPS287 & $3.32$ & $8.77_{-0.32}^{+0.30}$ & $7.47_{-0.48}^{+0.48}$ & $0.207_{-0.019}^{+0.020}$ & $0.37\pm0.21$ & $1.3$ \\
HPS288 & $3.04$ & $8.77_{-0.29}^{+0.28}$ & $7.31_{-0.37}^{+0.48}$ & $0.039_{-0.039}^{+0.010}$ & $1.35\pm0.04$ & $0.8$ \\
HPS292 & $2.87$ & $8.59_{-0.20}^{+0.19}$ & $7.71_{-0.29}^{+0.28}$ & $0.022_{-0.022}^{+0.006}$ & $0.73\pm0.04$ & $5.8$ \\
HPS296 & $2.84$ & $8.26_{-0.22}^{+0.21}$ & $7.51_{-0.31}^{+0.30}$ & $0.011_{-0.011}^{+0.002}$ & $0.97\pm0.11$ & $9.4$ \\
HPS306 & $2.44$ & $9.01_{-0.12}^{+0.12}$ & $7.73_{-0.20}^{+0.21}$ & $0.039_{-0.039}^{+0.009}$ & $1.35\pm0.04$ & $3.3$ \\
HPS310 & $3.07$ & $9.26_{-0.18}^{+0.22}$ & $8.54_{-0.25}^{+0.32}$ & $0.024_{-0.024}^{+0.004}$ & $0.87\pm0.05$ & $4.1$ \\
HPS313 & $2.10$ & $9.82_{-0.08}^{+0.08}$ & $7.74_{-0.12}^{+0.11}$ & $0.161_{-0.006}^{+0.007}$ & $2.34\pm0.02$ & $17.3$ \\
HPS315 & $3.07$ & $9.04_{-0.23}^{+0.23}$ & $7.42_{-0.34}^{+0.37}$ & $0.096_{-0.010}^{+0.011}$ & $3.03\pm0.09$ & $2.0$ \\
HPS316 & $2.81$ & $9.40_{-0.10}^{+0.11}$ & $8.28_{-0.18}^{+0.19}$ & $0.027_{-0.027}^{+0.006}$ & $1.26\pm0.03$ & $1.9$ \\
HPS318 & $2.46$ & $9.54_{-0.12}^{+0.12}$ & $7.79_{-0.23}^{+0.23}$ & $0.139_{-0.010}^{+0.011}$ & $2.42\pm0.07$ & $6.9$ \\
HPS338 & $2.60$ & $7.86_{-0.30}^{+0.43}$ & $6.87_{-1.05}^{+0.93}$ & $0.116_{-0.116}^{+0.014}$ & $3.52\pm0.13$ & $0.9$ \\
HPS341 & $2.93$ & $8.41_{-0.19}^{+0.18}$ & $6.91_{-0.55}^{+0.55}$ & $0.059_{-0.059}^{+0.017}$ & $2.53\pm0.50$ & $1.0$ \\
HPS360 & $2.92$ & $9.89_{-0.19}^{+0.18}$ & $7.96_{-0.53}^{+0.56}$ & $0.266_{-0.017}^{+0.046}$ & $1.61\pm0.04$ & $1.9$ \\
HPS370 & $3.18$ & $8.40_{-0.22}^{+0.22}$ & $7.38_{-0.32}^{+0.33}$ & $0.033_{-0.033}^{+0.008}$ & $0.94\pm0.02$ & $0.9$ \\
HPS372 & $2.76$ & $7.52_{-0.39}^{+0.44}$ & $7.03_{-1.18}^{+0.97}$ & $0.067_{-0.067}^{+0.017}$ & $0.36\pm0.06$ & $2.9$ \\
HPS389 & $2.59$ & $9.09_{-0.10}^{+0.10}$ & $7.72_{-0.17}^{+0.17}$ & $0.107_{-0.008}^{+0.010}$ & $1.28\pm0.02$ & $1.9$ \\
HPS391 & $2.96$ & $9.44_{-0.15}^{+0.13}$ & $8.13_{-0.33}^{+0.27}$ & $0.101_{-0.015}^{+0.016}$ & $0.96\pm0.01$ & $1.2$ \\
HPS395 & $2.27$ & $9.34_{-0.08}^{+0.09}$ & $8.36_{-0.17}^{+0.18}$ & $0.036_{-0.036}^{+0.009}$ & $3.79\pm0.52$ & $1.6$ \\
HPS402 & $2.97$ & $8.36_{-0.12}^{+0.14}$ & $6.97_{-0.21}^{+0.28}$ & $0.051_{-0.008}^{+0.008}$ & $1.87\pm0.03$ & $2.2$ \\
HPS403 & $3.18$ & $8.97_{-0.30}^{+0.28}$ & $8.17_{-0.46}^{+0.41}$ & $0.037_{-0.037}^{+0.010}$ & $2.70\pm0.16$ & $7.8$ \\
HPS415 & $3.37$ & $9.71_{-0.18}^{+0.15}$ & $9.05_{-0.21}^{+0.22}$ & $0.019_{-0.019}^{+0.004}$ & $1.09\pm0.03$ & $4.3$ \\
HPS419 & $2.23$ & $10.00_{-0.12}^{+0.13}$ & $8.83_{-0.25}^{+0.28}$ & $0.132_{-0.014}^{+0.019}$ & $1.67\pm0.02$ & $1.8$ \\
HPS420 & $2.93$ & $8.67_{-0.18}^{+0.18}$ & $7.71_{-0.28}^{+0.28}$ & $0.052_{-0.052}^{+0.011}$ & $0.56\pm0.01$ & $5.4$ \\
HPS426 & $3.40$ & $9.57_{-0.20}^{+0.20}$ & $8.84_{-0.30}^{+0.42}$ & $0.049_{-0.049}^{+0.011}$ & $0.70\pm0.02$ & $4.1$ \\
HPS428 & $3.34$ & $9.46_{-0.11}^{+0.10}$ & $8.04_{-0.20}^{+0.18}$ & $0.072_{-0.009}^{+0.010}$ & $1.93\pm0.07$ & $2.9$ \\
HPS434 & $2.27$ & $9.03_{-0.26}^{+0.31}$ & $8.74_{-0.56}^{+0.71}$ & $0.073_{-0.073}^{+0.026}$ & $1.03\pm0.04$ & $3.1$ \\
HPS436 & $2.42$ & $8.36_{-0.16}^{+0.15}$ & $7.23_{-0.26}^{+0.24}$ & $0.077_{-0.007}^{+0.009}$ & $1.78\pm0.02$ & $1.5$ \\
HPS447 & $3.13$ & $8.39_{-0.11}^{+0.10}$ & $6.36_{-0.74}^{+0.60}$ & $0.033_{-0.033}^{+0.009}$ & $3.60\pm0.11$ & $2.5$ \\
HPS462 & $2.21$ & $10.48_{-0.11}^{+0.13}$ & $8.61_{-0.29}^{+0.32}$ & $0.296_{-0.020}^{+0.024}$ & $1.38\pm0.84$ & $2.8$ \\
HPS466 & $3.24$ & $9.04_{-0.03}^{+0.03}$ & $6.82_{-0.04}^{+0.05}$ & $0.134_{-0.004}^{+0.004}$ & $1.51\pm0.01$ & $12.9$ \\
HPS474 & $2.27$ & $8.97_{-0.20}^{+0.20}$ & $7.95_{-0.33}^{+0.35}$ & $0.096_{-0.010}^{+0.014}$ & $1.18\pm0.02$ & $8.3$ \\
\enddata
\end{deluxetable}

\end{document}